\documentclass[preprint,12pt]{elsarticle}

\usepackage{amssymb}
\journal{}

\begin{document}

\begin{frontmatter}



\title{Half-metallic ferromagnetism in the RbSe and CsTe compounds with CsCl structure: A first-principles study
}

\author[mymainaddress]{Huan-Huan Xie}
\author[mymainaddress]{Qiang Gao}
\author[mymainaddress]{Lei Li}
\author[mymainaddress]{Gang Lei}
\author[mymainaddress]{Xian-Ru Hu}

\author[mymainaddress]{Jian-Bo Deng\corref{mycorrespondingauthor}}
\cortext[mycorrespondingauthor]{Corresponding author}
\ead{dengjb@lzu.edu.cn}

\address[mymainaddress]{School of Physical Science and Technology, Lanzhou University,
 Lanzhou 730000, People's Republic of China}

\begin{abstract}
We investigate the electronic structures and magnetic properties of RbSe and CsTe compounds in CsCl, RS and ZB structures  by using first-principles calculation. It is shown that these two  compounds exhibit half-metallic ferromagnetism with an integer magnetic moment of 1.00 $\mu_B$ per formula in all the three structures. Total energy calculations indicate the CsCl phase is more stable than the other two phases. We investigate these two compounds with CsCl structure in detail. The ferromagnetism results mainly from the spin-polarization of p states of anion Se (Te) for the two compounds. The half-metallicities of RbSe and CsTe compounds can be preserved up to 2.5\% and 0.8\% compression of lattice constants with respect to their equilibrium lattices, respectively. 
\end{abstract}


\end{frontmatter}

% \linenumbers

\section{Introduction}
\label{}
The new and rapidly developing field of spintronics, in which the spin degree of freedom of the electrons is used in addition to the charge of the electrons, has provided strong motivation to search novel half-metallic  materials \cite{1}.

Half-metallic magnets,  are seen as the most promising candidates of high-spin-polarization materials, because their band structure is metallic in one of the two spin channels and semiconducting or insulating in the other one, which results in complete (100\%) spin polarization of electrons at the Fermi level. 

In 1983, de Groot et al. \cite{2} predicted the first half-metallic ferromagnet of the half-Heusler alloy NiMnSb. Since then, a number of new half-metallic materials, such as CrAs, NiMnSb, Co$_2$MnAl, etc. \cite{3-1,3-2}, have been initially predicted theoretically by first-principle calculations and later verified by experiments. 

Half-metallic properties are present in various perovskite structures \cite{4-1,4-2}, Heusler compounds \cite{5-1,5-2,5-3,5-4}, diluted magnetic semi-conductors \cite{6} and metallic oxides \cite{7-1,7-2}. All these materials contain transition metal atoms. Recently, calcium pnictide compounds such as CaP, CaAs, and CaSb with zinc-blende structure were reported to be half-metallic ferromagnets with magnetic moment of 1.0 $\mu_B$. These pnictides do not include any transition-metal atoms and ferromagnetism does not arise from the p-d exchange and the double exchange which are significant in 3d transition-metal compounds \cite{8}. Here the spin polarization of the s and p states makes main contribution to the half-metallic ferromgnetism. This kind of materials is then considered as the sp half-metallic ferromagnets. It is well known that the ferromagnets with large magnetic moment wpuld give rise to more stray flux, which can lead to more energy losses in devices \cite{9}. So sp half-metallic ferromagnets such as CaC and CaN are more meaningful, because they possess small values of magnetic moment (1 or 2 $\mu_B$ per formula unit \cite{10-1,10-2}) compared to those of d half-metallic ferromagnets  such as CrAs and CrTe (3 or 4 $\mu_B$ per formula unit \cite{11}). Later, alkaline metal pnictides wuth rocksalt structure \cite{13} and alkaline metal LiS, NaS and KS \cite{14} were reported to be half-metallic ferromagnets.

These studies motivated us to search for new sp half-metallic materials which do not contain 3d transition metals. To the best of our knowledge, there is no experimental and theoretical study about the RbSe and CsTe compounds. In the present work, we report the theoretical investigation of the electronic structure and magnetism for ZB, RS and CsCl structures about the two alloys. We predict that the two alloys are HM ferromagnets with an magnetic moment of 1.0 $\mu_B$ in all the three structures. The CsCl structure are energetically the most stable structure for the two alloys.

\section{Method of calculations}
\label{}
We have carried out density functional calculation using the scalar relativistic version of the full-potential local-orbital (FPLO) minimum-basis band-structure method \cite{15-1,15-2}. For the present calculations, the site-centered potentials and densities are expanded in spherical harmonic contributions up to $l$$_{max}$=12. The Perdew-Burke-Ernzerhof 96 of the generalized gradient approximation (GGA) is used for exchange-correlation (XC) potential \cite{16}. For the irreducible  brillouin zone, we use the $k$ meshes of 20$\times$20$\times$20 for all the calculations. The convergence criteria of self-consistent iterations is set to $10^{-6}$ to the density and $10^{-8}$ Hartree to the total energy per formula unit.

\section{Results and discussions}
\label{}
CsCl, RS and ZB phases are three different crystal structures. CsCl structure has the space group Pm$\bar{3}$m (no. 221); RS has the space group Fm$\bar{3}$m (no. 225); ZB has the space group F$\bar{4}$3m (no. 216).  Rb(Cs) and Se(Te) are located at (0,0,0) and (0.5,0.5,0.5); (0,0,0) and (0.5,0.5,0.5); (0,0,0) and (0.25,0.25,0.25) of these two compounds in the three structures, respectively. We calculate the ferromagnetic (FM) and paramagnetic (PM) states total energies as a function of lattice constants for RbSe and CsTe in the three structures. The results are shown in Figure. 1. Note that the total energy in the PM state is the relative value to that in the FM state. Moreover, the FM state is more stable than the PM state. From Fig. 1, we find that the FM CsCl structures are energetically the most atable structures for both RbSe and CsTe compounds. 

Table 1 presents the optimized lattice constants and the calculated total magnetic moments per formula unit at equilibrium lattices for the two compounds in the three structures. Half-metallic gaps, the formation and cohesive energies are also listed in Table 1. 

The formation energy (E$_f$) and cohesive energy (E$_c$) indicating thermal stability of a compound can be calculated using the following formulas:
\begin{equation}
E_{f}=E^{tot}_{RbSe}-[E^{bulk}_{Rb}+E^{bulk}_{Se}] 
\end{equation}  

\begin{equation}
E_{c}=E^{tot}_{RbSe}-[E^{iso}_{Rb}+E^{iso}_{Se}] 
\end{equation}  
where $E^{tot}_{RbSe}$ is the total energy of the RbSe compound, $E^{bulk}_{Rb}$ and $E^{bulk}_{Se}$ refers to the total energy per atom of each elemental bulk. $E^{iso}_{Rb}$ and $E^{iso}_{Se}$ are the energies of an isolated Cs and Se atoms, respectively. It is same for the CsTe compound, just replace Rb or Se by Cs or Te in the relevant position. The negative values of the formation energy and cohesive energy, listed in Table 1 for this two alloys in the three structures, indicate that these two compounds are energetically stable.

The calculated spin-polarized total densities of states (DOS) are plotted in Figure. 2 for the RbSe and CsTe in the three structures at their optimized lattice parameters. The total DOS of the RbSe and CsTe compounds in the same types are generally similar in shape. It is clear that there is an energy gap at Fermi level in the spin-up states, while the spin-down states are metallic. The energy gap seen in the spin-up band causes 100\% spin polarization at Fermi level and half-metallic ferromagnetism. Non-zero spin flip gap proves that these two compounds with the three structures are ture half-metallic ferromagnets. 

Since the CsCl structure is energetically the most stable one among the three structures, next we will discuss the properties of RbSe and CsTe in this structure. The values of the valence band maximum (VBM), conduction band minimum (CBM) and band gap are given in Table 2 for the two compounds in CsCl structure. The energy required to create a spin hole at the top of the spin-up valence bands by exciting a spin-up valence band electron into the spin-down conducting bands means the spin flip gap which is equal to the absolute value of the valence band maximum. Figure. 3 shows the calculated results of total and partial DOS of RbSe and CsTe in CsCl structure. As for RbSe, the energy region below  -8.0 eV consists mainly of Rb-p and Se-s electrons. The density of states near the Fermi level are originated mainly from the Se-p orbitals, with relatively small contribution of the Rb-s and Rb-d states. In the case of CsTe, the energy region below  -6.0 eV consists mainly of Cs-p and Te-s electrons. The density of states near the Fermi level are originated mainly from the Te-p orbitals, with relatively small contribution of the Cs-s and Cs-d states. The calculated values of the total and partial magnetic moment of these two compounds in CsCl structure are listed in Table. 2. The values of the total magnetic moments are evaluated to be 1.0 $\mu_B$ for the two compounds. The formula $M_{tot}$= n - 8 can be used to calculate the total magnetic moment, where n is the sum of valence electrons per formula unit. For the two compounds, the main contribution to the total magnetic moment comes from the Se/Te atom, while the contribution of Rb/Cs atom is very small.

The band structure of RbSe and CsTe compounds at their optimized lattice constant for the spin-up and spin-down electrons is plotted in Figure. 4 and Figure. 5. Definitely, these two compounds exhibit HM characteristics: the spin-down band structure is metallic, and there is an energy gap in the spin-up band structure. The energy gap can be determined from the difference between the energies of the lowest unoccupied band at the $\Gamma$ point and the highest occupied band at the M point.

In order to study the stability of the half-metallicity along with the variation of the interatomic distances, we further calculated the total magnetic moment with the contraction of lattice constant (see Fgure. 6). From Figure. 6, we can see that the total magnetic moments keep an integer value of 1.0 $\mu_B$ per formula unit the lattice constants are contracted to be 4.063 and 4.585 \AA for RbSe and CsTe, respectively. The half-metallicities of RbSe and CsTe can be preserved up to 2.5\% and 0.8\% compression of lattice constants with respect to their equilibrium lattices, respectively. In addition, we also reveal that RbSe and CsTe are still HM when their lattice constants are expanded appropriately. The changes of magnetic moments for RbSe and CsTe compounds under pressure are given in Figure. 7. As can be seen from it, the magnetic moments keeps 1.0 $\mu_B$ when pressure in the range from 0 to 2.12 GPa and 0 to 0.29 GPa for RbSe and CsTe compounds, respectively.

\section{Conclusions}
\label{}
In conclusion, we have used first-principle FPLO method to investigate the electronic  structures and magnetic properties of RbSe and CsTe compounds with CsCl, RS and ZB structures. This two compounds exhibit the half-metallic character with magnetic moment of 1.0 $\mu_B$ in all the three structures. These ferromagnetic compounds do not include and 3d transition metal element and the magnetism mainly arises from the spin polarization of the Se-p/Te-p states. The compounds with CsCl structure are energetically more stable than the other two structures, which makes them more promising materials for future spintronic applications.



%% References

\newpage
\bibliographystyle{elsarticle-num}
\bibliography{Reference.bib}

\begin{thebibliography}{10}
\expandafter\ifx\csname url\endcsname\relax
  \def\url#1{\texttt{#1}}\fi
\expandafter\ifx\csname urlprefix\endcsname\relax\def\urlprefix{URL }\fi
\expandafter\ifx\csname href\endcsname\relax
  \def\href#1#2{#2} \def\path#1{#1}\fi

\bibitem{1}
G.~A. Prinz,
  \href{http://www.sciencemag.org/content/282/5394/1660}{Magnetoelectronics},
  Science 282~(5394) (1998) 1660--1663.
\newblock \href {http://dx.doi.org/10.1126/science.282.5394.1660}
  {\path{doi:10.1126/science.282.5394.1660}}.
\newline\urlprefix\url{http://www.sciencemag.org/content/282/5394/1660}

\bibitem{2}
R.~A. de~Groot, F.~M. Mueller, P.~G.~v. Engen, K.~H.~J. Buschow,
  \href{http://link.aps.org/doi/10.1103/PhysRevLett.50.2024}{New {Class of
  Materials}: Half-metallic ferromagnets}, Physical Review Letters 50~(25)
  (1983) 2024--2027.
\newblock \href {http://dx.doi.org/10.1103/PhysRevLett.50.2024}
  {\path{doi:10.1103/PhysRevLett.50.2024}}.
\newline\urlprefix\url{http://link.aps.org/doi/10.1103/PhysRevLett.50.2024}

\bibitem{3-1}
I.~Galanakis,
  \href{http://openurl.ingenta.com/content/xref?genre=article,issn=2164-7542,volume=2,issue=1,spage=74}{Slater-pauling
  {Behavior in Half-Metallic Magnets}}, Journal of Surfaces and Interfaces of
  Materials 2~(1) (2014) 74--78.
\newblock \href {http://dx.doi.org/10.1166/jsim.2014.1037}
  {\path{doi:10.1166/jsim.2014.1037}}.
\newline\urlprefix\url{http://openurl.ingenta.com/content/xref?genre=article,issn=2164-7542,volume=2,issue=1,spage=74}

\bibitem{3-2}
A.~Hirohata, M.~Kikuchi, N.~Tezuka, K.~Inomata, J.~S. Claydon, Y.~B. Xu,
  G.~van~der Laan,
  \href{http://www.sciencedirect.com/science/article/pii/S1359028606001124}{Heusler
  alloy/semiconductor hybrid structures}, Current Opinion in Solid State and
  Materials Science 10~(2) (2006) 93--107.
\newblock \href {http://dx.doi.org/10.1016/j.cossms.2006.11.006}
  {\path{doi:10.1016/j.cossms.2006.11.006}}.
\newline\urlprefix\url{http://www.sciencedirect.com/science/article/pii/S1359028606001124}

\bibitem{4-1}
Y.~P. Liu, H.~R. Fuh, Z.~R. Xiao, Y.~K. Wang,
  \href{http://www.sciencedirect.com/science/article/pii/S0925838813024341}{Theoretical
  prediction of half-metallic materials in double perovskites
  {Sr}$_2${Cr(Co)B′O}$_6$ {(B′ = Y, La, Zr, and Hf)} and
  {Sr}$_2${V(Fe)B′O}$_6$ {(B′ = Zr and Hf)}}, Journal of Alloys and
  Compounds 586 (2014) 289--294.
\newblock \href {http://dx.doi.org/10.1016/j.jallcom.2013.10.043}
  {\path{doi:10.1016/j.jallcom.2013.10.043}}.
\newline\urlprefix\url{http://www.sciencedirect.com/science/article/pii/S0925838813024341}

\bibitem{4-2}
B.~Sahli, H.~Bouafia, B.~Abidri, A.~Abdellaoui, S.~Hiadsi, A.~Akriche,
  N.~Benkhettou, D.~Rached,
  \href{http://www.sciencedirect.com/science/article/pii/S0925838815005587}{First-principles
  prediction of structural, elastic, electronic and thermodynamic properties of
  the cubic {SrUO}$_3$-perovskite}, Journal of Alloys and Compounds 635 (2015)
  163--172.
\newblock \href {http://dx.doi.org/10.1016/j.jallcom.2015.02.118}
  {\path{doi:10.1016/j.jallcom.2015.02.118}}.
\newline\urlprefix\url{http://www.sciencedirect.com/science/article/pii/S0925838815005587}

\bibitem{5-1}
N.~Zheng, Y.~Jin,
  \href{http://www.sciencedirect.com/science/article/pii/S0304885312004155}{Band-gap
  and {Slater–Pauling} rule in half-metallic {Ti}$_2$-based {Heusler} alloys:
  A first-principles study}, Journal of Magnetism and Magnetic Materials
  324~(19) (2012) 3099--3104.
\newblock \href {http://dx.doi.org/10.1016/j.jmmm.2012.05.009}
  {\path{doi:10.1016/j.jmmm.2012.05.009}}.
\newline\urlprefix\url{http://www.sciencedirect.com/science/article/pii/S0304885312004155}

\bibitem{5-2}
L.~Guan-Nan, J.~Ying-Jiu, L.~J. Il,
  \href{http://iopscience.iop.org/1674-1056/19/9/097102}{Effect of local atomic
  disorder on the half-metallicity of full-heusler {Co}$_2${FeSi} alloy: a
  first-principles study}, Chinese Physics B 19~(9) (2010) 097102.
\newblock \href {http://dx.doi.org/10.1088/1674-1056/19/9/097102}
  {\path{doi:10.1088/1674-1056/19/9/097102}}.
\newline\urlprefix\url{http://iopscience.iop.org/1674-1056/19/9/097102}

\bibitem{5-3}
J.~Chen, G.~Y. Gao, K.~L. Yao, M.~H. Song,
  \href{http://www.sciencedirect.com/science/article/pii/S0925838811017701}{Half-metallic
  ferromagnetism in the {half-Heusler compounds GeKCa and SnKCa from}
  first-principles calculations}, Journal of Alloys and Compounds 509~(42)
  (2011) 10172--10178.
\newblock \href {http://dx.doi.org/10.1016/j.jallcom.2011.08.083}
  {\path{doi:10.1016/j.jallcom.2011.08.083}}.
\newline\urlprefix\url{http://www.sciencedirect.com/science/article/pii/S0925838811017701}

\bibitem{5-4}
G.~Y. Gao, L.~Hu, K.~L. Yao, B.~Luo, N.~Liu,
  \href{http://www.sciencedirect.com/science/article/pii/S0925838812020610}{Large
  half-metallic gaps in the quaternary heusler alloys {CoFeCrZ (Z = Al, Si,
  Ga, Ge): A first-principles study}}, Journal of Alloys and Compounds 551
  (2013) 539--543.
\newblock \href {http://dx.doi.org/10.1016/j.jallcom.2012.11.077}
  {\path{doi:10.1016/j.jallcom.2012.11.077}}.
\newline\urlprefix\url{http://www.sciencedirect.com/science/article/pii/S0925838812020610}

\bibitem{6}
Z.~F. Wu, K.~Cheng, F.~Zhang, R.~F. Guan, X.~M. Wu, L.~J. Zhuge,
  \href{http://www.sciencedirect.com/science/article/pii/S0925838814015746}{Effect
  of {Al} co-doping on the electrical and magnetic properties of {Cu-doped ZnO}
  nanorods}, Journal of Alloys and Compounds 615 (2014) 521--525.
\newblock \href {http://dx.doi.org/10.1016/j.jallcom.2014.06.204}
  {\path{doi:10.1016/j.jallcom.2014.06.204}}.
\newline\urlprefix\url{http://www.sciencedirect.com/science/article/pii/S0925838814015746}

\bibitem{7-1}
M.~Karaca, S.~Kervan, N.~Kervan,
  \href{http://www.sciencedirect.com/science/article/pii/S0925838815008865}{Half-metallic
  ferromagnetism in the {CsSe} compound by density functional theory}, Journal
  of Alloys and Compounds 639 (2015) 162--167.
\newblock \href {http://dx.doi.org/10.1016/j.jallcom.2015.03.164}
  {\path{doi:10.1016/j.jallcom.2015.03.164}}.
\newline\urlprefix\url{http://www.sciencedirect.com/science/article/pii/S0925838815008865}

\bibitem{7-2}
Y.~S. Dedkov, U.~Rüdiger, G.~Güntherodt,
  \href{http://link.aps.org/doi/10.1103/PhysRevB.65.064417}{Evidence for the
  half-metallic ferromagnetic state of {Fe}$_3${O}$_4$ by spin-resolved
  photoelectron spectroscopy}, Physical Review B 65~(6) (2002) 064417.
\newblock \href {http://dx.doi.org/10.1103/PhysRevB.65.064417}
  {\path{doi:10.1103/PhysRevB.65.064417}}.
\newline\urlprefix\url{http://link.aps.org/doi/10.1103/PhysRevB.65.064417}

\bibitem{8}
K.~Kusakabe, M.~Geshi, H.~Tsukamoto, N.~Suzuki,
  \href{http://iopscience.iop.org/0953-8984/16/48/021}{New half-metallic
  materials with an alkaline earth element}, Journal of Physics: Condensed
  Matter 16~(48) (2004) S5639.
\newblock \href {http://dx.doi.org/10.1088/0953-8984/16/48/021}
  {\path{doi:10.1088/0953-8984/16/48/021}}.
\newline\urlprefix\url{http://iopscience.iop.org/0953-8984/16/48/021}

\bibitem{9}
I.~Galanakis, K.~Özdoğan, E.~Şaşıoğlu, B.~Aktaş,
  \href{http://link.aps.org/doi/10.1103/PhysRevB.75.092407}{Doping of
  {Mn}$_2${VAl} and {Mn}$_2${VSi} {Heusler} alloys as a route to half-metallic
  antiferromagnetism}, Physical Review B 75~(9) (2007) 092407.
\newblock \href {http://dx.doi.org/10.1103/PhysRevB.75.092407}
  {\path{doi:10.1103/PhysRevB.75.092407}}.
\newline\urlprefix\url{http://link.aps.org/doi/10.1103/PhysRevB.75.092407}

\bibitem{10-1}
O.~Volnianska, P.~Jakubas, P.~Bogusławski,
  \href{http://www.sciencedirect.com/science/article/pii/S0925838806002179}{Magnetism
  of caas, cap and can half-metals}, Journal of Alloys and Compounds
  423~(1–2) (2006) 191--193.
\newblock \href {http://dx.doi.org/10.1016/j.jallcom.2006.01.092}
  {\path{doi:10.1016/j.jallcom.2006.01.092}}.
\newline\urlprefix\url{http://www.sciencedirect.com/science/article/pii/S0925838806002179}

\bibitem{10-2}
G.~Y. Gao, K.~L. Yao, E.~Şaşıoğlu, L.~M. Sandratskii, Z.~L. Liu, J.~L.
  Jiang,
  \href{http://link.aps.org/doi/10.1103/PhysRevB.75.174442}{Half-metallic
  ferromagnetism in zinc-blende {CaC, SrC, and BaC} from first principles},
  Physical Review B 75~(17) (2007) 174442.
\newblock \href {http://dx.doi.org/10.1103/PhysRevB.75.174442}
  {\path{doi:10.1103/PhysRevB.75.174442}}.
\newline\urlprefix\url{http://link.aps.org/doi/10.1103/PhysRevB.75.174442}

\bibitem{11}
I.~Galanakis, P.~Mavropoulos,
  \href{http://link.aps.org/doi/10.1103/PhysRevB.67.104417}{Zinc-blende
  compounds of transition elements with {N, P, As, Sb, S, Se and Te} as
  half-metallic systems}, Physical Review B 67~(10) (2003) 104417.
\newblock \href {http://dx.doi.org/10.1103/PhysRevB.67.104417}
  {\path{doi:10.1103/PhysRevB.67.104417}}.
\newline\urlprefix\url{http://link.aps.org/doi/10.1103/PhysRevB.67.104417}

\bibitem{13}
G.~Y. Gao, K.~L. Yao, Z.~L. Liu, Y.~Min, J.~Zhang, S.~W. Fan, D.~H. Zhang,
  \href{http://iopscience.iop.org/0953-8984/21/27/275502}{Bulk and surface sp
  half-metallic ferromagnetism in alkali metal pnictides with rocksalt
  structure: a first-principles calculation}, Journal of Physics: Condensed
  Matter 21~(27) (2009) 275502.
\newblock \href {http://dx.doi.org/10.1088/0953-8984/21/27/275502}
  {\path{doi:10.1088/0953-8984/21/27/275502}}.
\newline\urlprefix\url{http://iopscience.iop.org/0953-8984/21/27/275502}

\bibitem{14}
G.~Y. Gao, K.~L. Yao, M.~H. Song, Z.~L. Liu,
  \href{http://www.sciencedirect.com/science/article/pii/S0304885311003350}{Half-metallic
  ferromagnetism in rocksalt and zinc-blende {MS (M=Li, Na and K): A
  first-principles study}}, Journal of Magnetism and Magnetic Materials
  323~(21) (2011) 2652--2657.
\newblock \href {http://dx.doi.org/10.1016/j.jmmm.2011.06.003}
  {\path{doi:10.1016/j.jmmm.2011.06.003}}.
\newline\urlprefix\url{http://www.sciencedirect.com/science/article/pii/S0304885311003350}

\bibitem{15-1}
I.~Opahle, K.~Koepernik, H.~Eschrig,
  \href{http://link.aps.org/doi/10.1103/PhysRevB.60.14035}{Full-potential
  band-structure calculation of iron pyrite}, Physical Review B 60~(20) (1999)
  14035--14041.
\newblock \href {http://dx.doi.org/10.1103/PhysRevB.60.14035}
  {\path{doi:10.1103/PhysRevB.60.14035}}.
\newline\urlprefix\url{http://link.aps.org/doi/10.1103/PhysRevB.60.14035}

\bibitem{15-2}
K.~Koepernik, H.~Eschrig,
  \href{http://link.aps.org/doi/10.1103/PhysRevB.59.1743}{Full-potential
  nonorthogonal local-orbital minimum-basis band-structure scheme}, Physical
  Review B 59~(3) (1999) 1743--1757.
\newblock \href {http://dx.doi.org/10.1103/PhysRevB.59.1743}
  {\path{doi:10.1103/PhysRevB.59.1743}}.
\newline\urlprefix\url{http://link.aps.org/doi/10.1103/PhysRevB.59.1743}

\bibitem{16}
J.~P. Perdew, K.~Burke, M.~Ernzerhof,
  \href{http://link.aps.org/doi/10.1103/PhysRevLett.77.3865}{Generalized
  gradient approximation made simple}, Physical Review Letters 77~(18) (1996)
  3865--3868.
\newblock \href {http://dx.doi.org/10.1103/PhysRevLett.77.3865}
  {\path{doi:10.1103/PhysRevLett.77.3865}}.
\newline\urlprefix\url{http://link.aps.org/doi/10.1103/PhysRevLett.77.3865}

\end{thebibliography}






% \end{thebibliography}

\newpage
Table captions: \\

Figure captions: \\

\newpage
\begin{table}[!hbp]

\caption{The calculated equilibrium lattice constants, magnetic moments, half-metallic gap ($E_g$), formation (E$_f$) and cohesive (E$_c$) energise for the RbSe and CsTe compounds with the three different structures.}
\footnotesize\rm
\begin{tabular*}{\textwidth}{@{}l*{15}{@{\extracolsep{0pt plus12pt}}l}}
\hline
\hline

Compounds&a (\AA)&m ($\mu_B$)&E$_g$ (eV)& E$_f$ (eV) & E$_c$ (eV)\\

\hline
RbSe-CsCl & 4.167 & 1.00 & 0.06 & -1.92 & -5.50\\
RbSe-RS   & 7.087 & 1.00 & 0.30 & -1.68 & -5.26\\
RbSe-ZB   & 7.908 & 1.00 & 0.57 & -1.39 & -4.97\\
CsTe-CsCl & 4.621 & 1.00 & 0.02 & -1.61 & -5.02\\
CsTe-RS   & 7.874 & 1.00 & 0.21 & -1.37 & -4.77\\
CsTe-ZB   & 8.794 & 1.00 & 0.41 & -1.10 & -4.50\\

\hline
\hline

\end{tabular*}
\end{table}

\newpage
\begin{table}[!hbp]
\caption{Valence band maximum (VBM), conduction band minimum (CBM), band gap and the total and partial magnetic moments for these two compounds in CsCl structure. Rb(Cs) and Se(Te) are represented by the number 1 and 2.}
\begin{tabular*}{\textwidth}{@{}l*{15}{@{\extracolsep{0pt plus12pt}}l}}
\hline
\hline
	compounds & VBM & CBM & Band gap&   $M_{tot}$ & $M_{1}$ & $M_{2}$ & \\
	          & (eV) & (eV)  & (eV) & ($\mu_B$) &  ($\mu_B$)   & ($\mu_B$)   &\\
\hline
	RbSe & -0.06 & 3.40 & 3.46 & 1.00 & -0.011 & 1.011\\
    CsTe & -0.02 & 3.28 & 3.30 & 1.00 & -0.010 & 1.010\\
	\hline
	\hline
\end{tabular*}
\end{table}

\newpage
\begin{figure}[htp]
\centering
\includegraphics[scale=0.7]{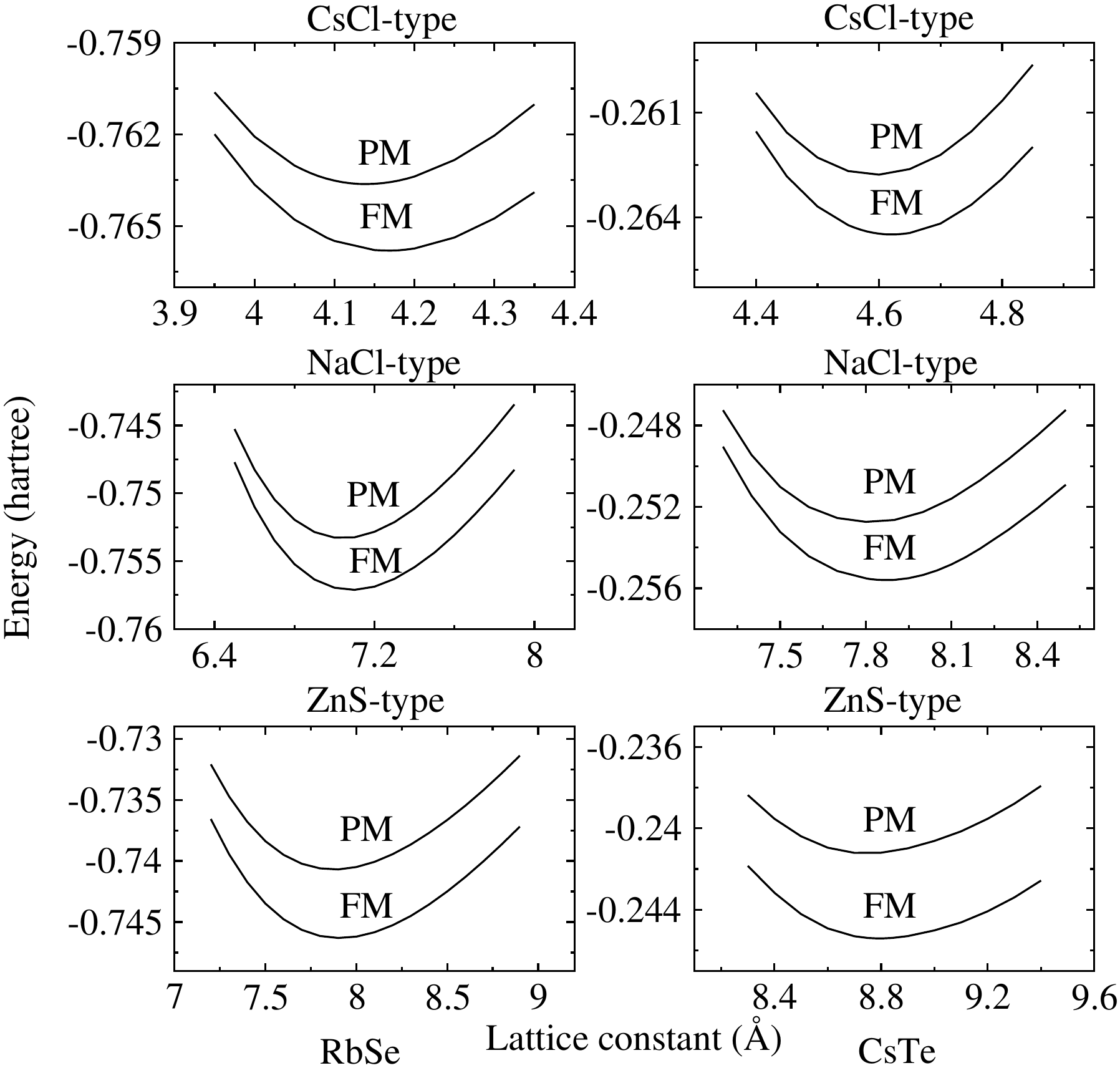}
\caption{Total energy as a function of lattice constant of RbSe and CsTe compounds in the three structures. PM and FM correspond to paramagnetic and ferromagnetic calculations, respectively.}
\label{}
\end{figure}

\newpage
\begin{figure}[htp]
\centering
\includegraphics[scale=0.6,angle=270]{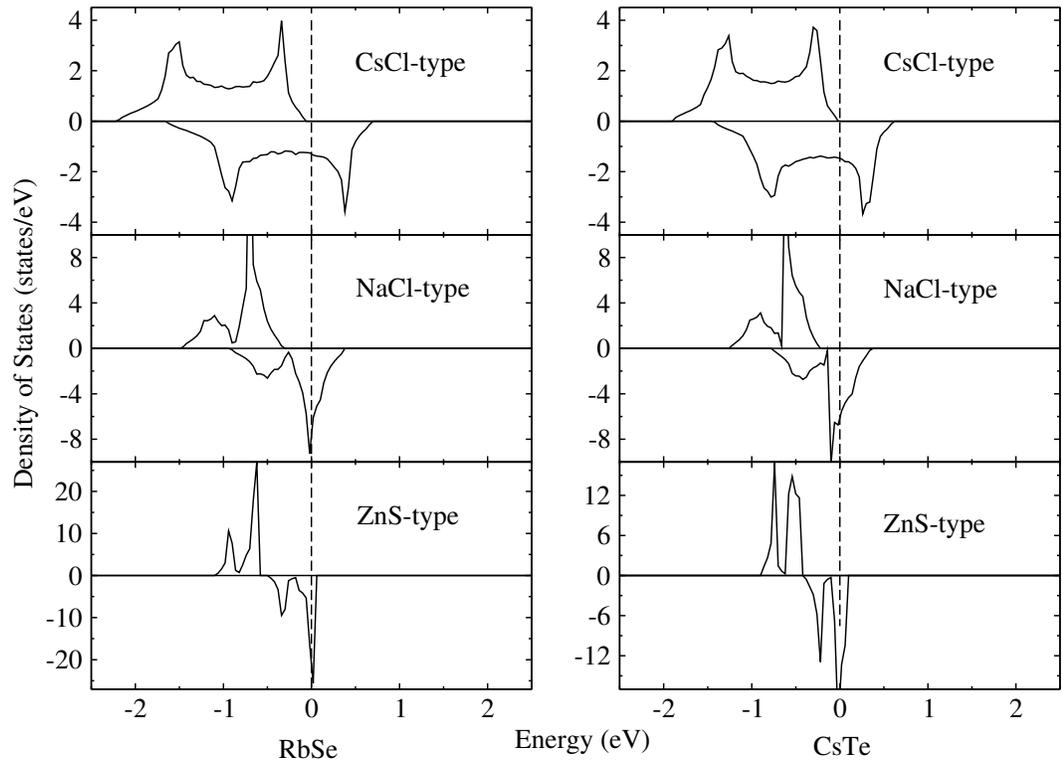}
\caption{ The total density of states for RbSe and CsTe compounds in the three structures at their equilibrium lattice constants. }
\label{}
\end{figure}

\newpage
\begin{figure}[htp]
\center
\includegraphics[scale=0.7]{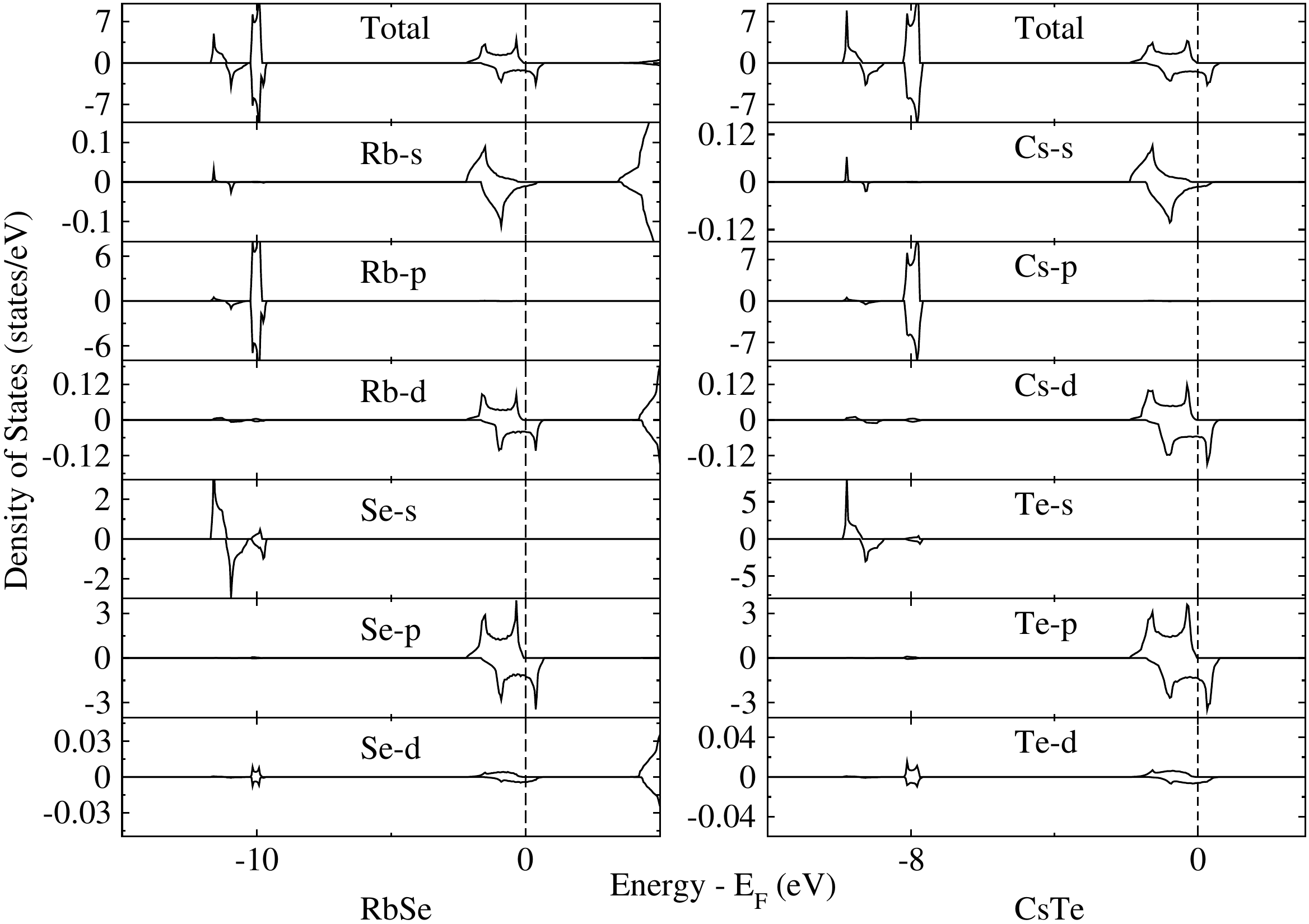}
\caption{ The total and partial DOS plots  of RbSe and CsTe at their equilibrium lattice constants. The zero energy value is in correspondence to the Fermi level. Positive values of DOS represent spin-up electrons, negative values represent spin-down electrons.}
\label{}
\end{figure}

\newpage
\begin{figure}[htp]
\center
\includegraphics[scale=0.6,angle=270]{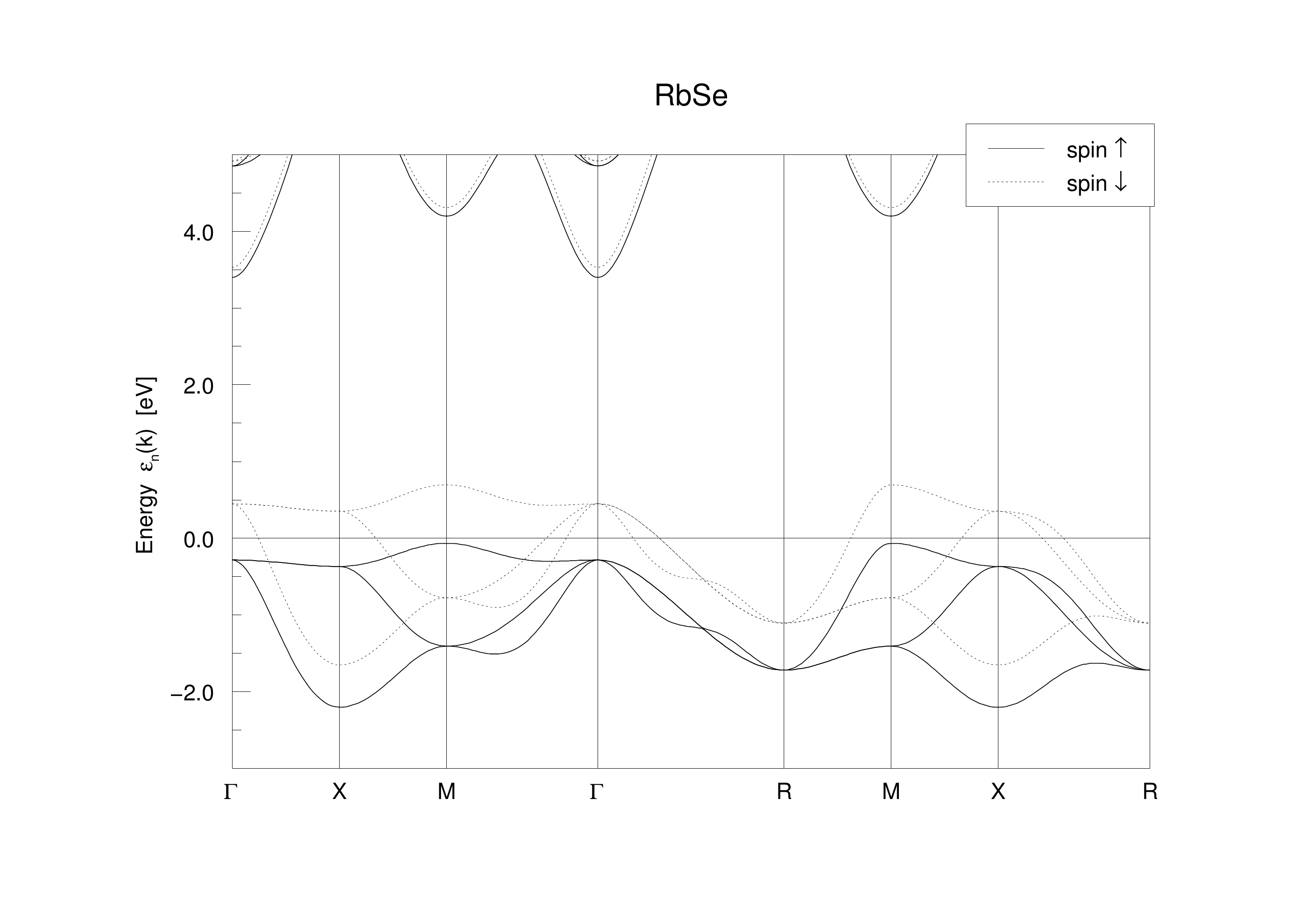}
\caption{ Spin-up and spin-down band structure for RbSe at the equilibrium lattice constant.  Arrows $\uparrow$ and $\downarrow$ represent the spin-up and spin-down states respectively.}
\label{}
\end{figure}

\newpage
\begin{figure}[htp]
\center
\includegraphics[scale=0.6,angle=270]{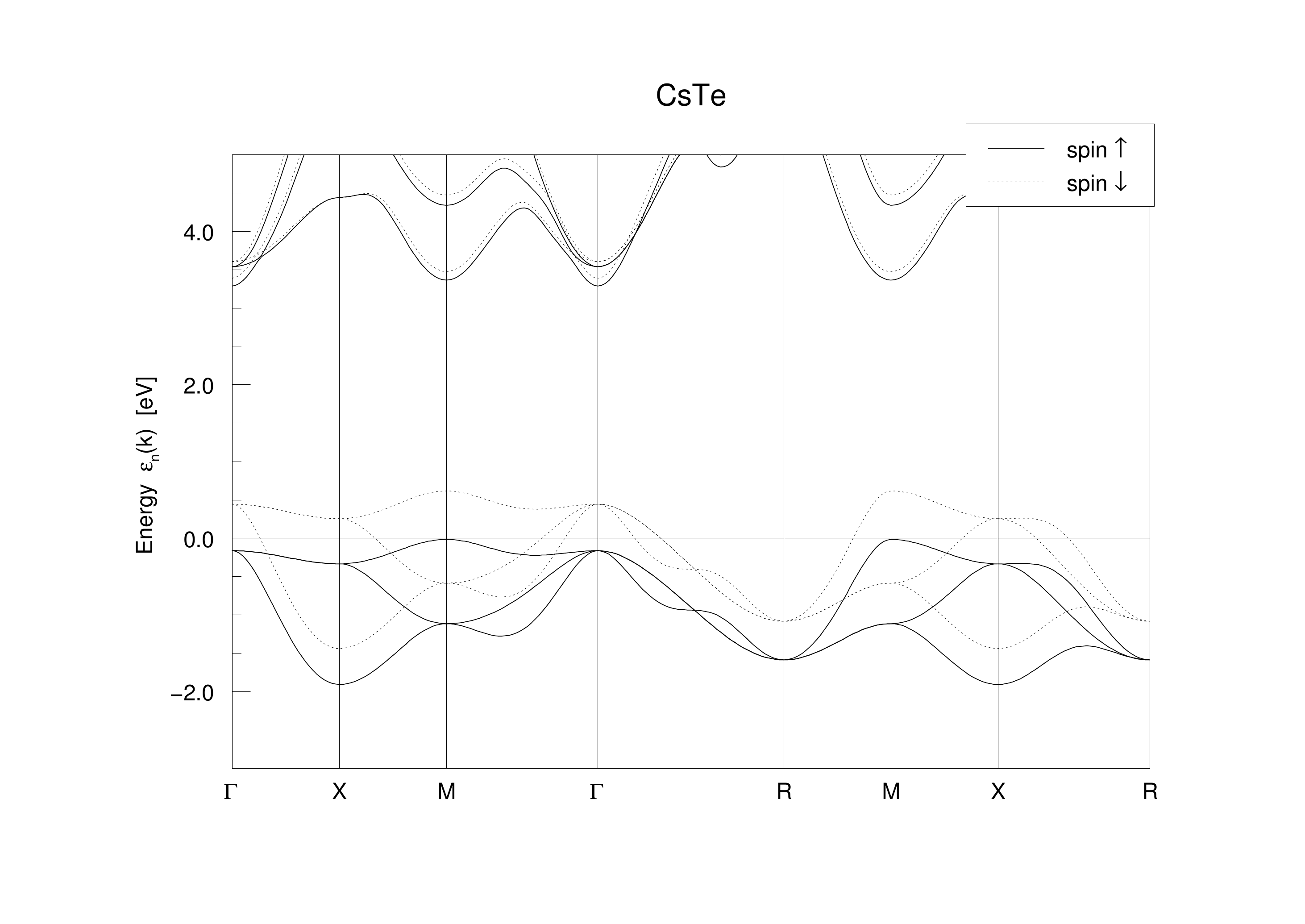}
\caption{ Spin-up and spin-down band structure for CsTe at the equilibrium lattice constant. Arrows $\uparrow$ and $\downarrow$ represent the spin-up and spin-down states respectively.}
\label{}
\end{figure}

\newpage
\begin{figure}[htp]
\center
\includegraphics[scale=0.6]{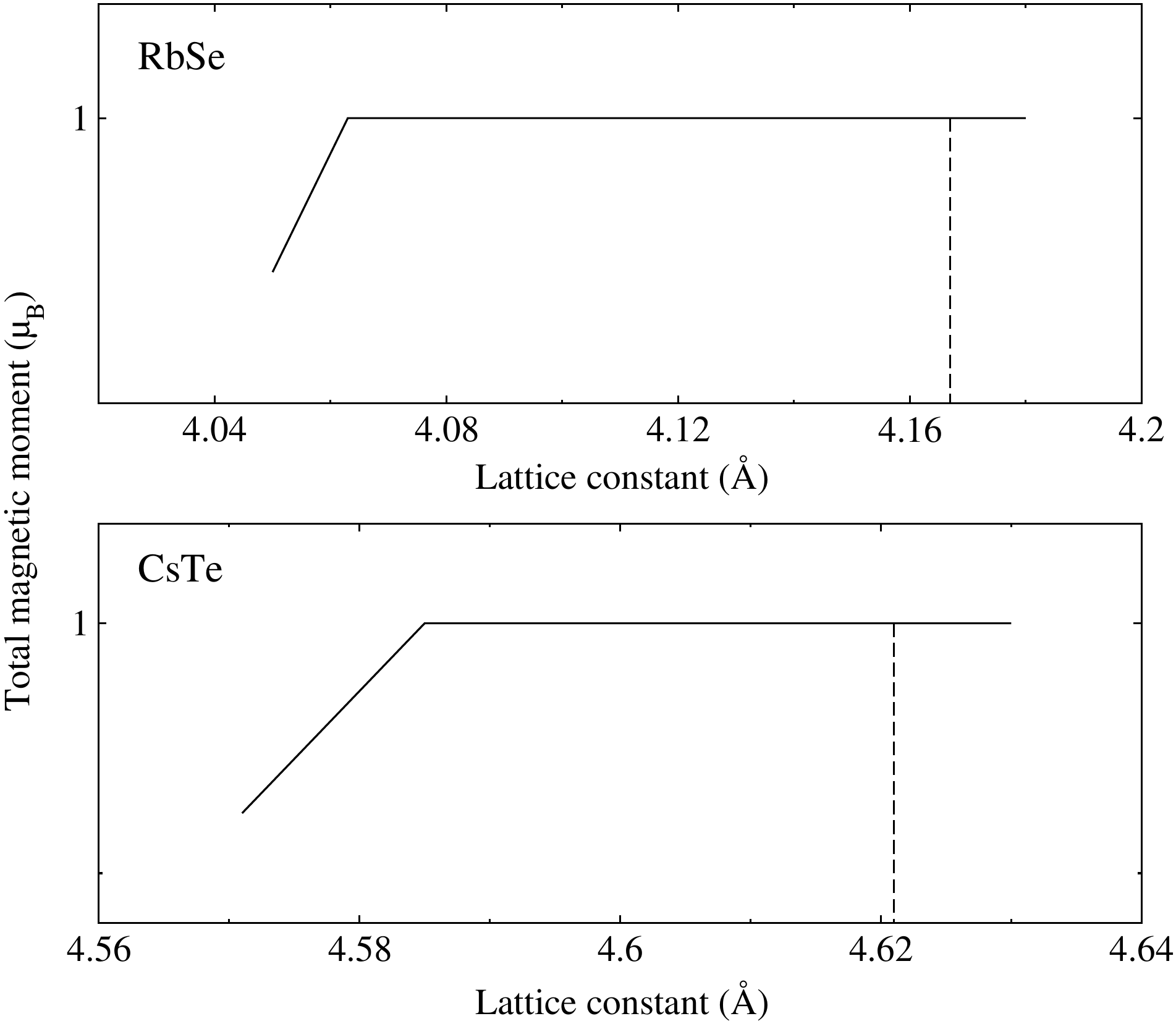}
\caption{ Total magnetic moments per formula unit  as a function of lattice constants for RbSe and CsTe. The dashed lines indicate the equilibrium lattice constants. }
\label{}
\end{figure}

\newpage
\begin{figure}[htp]
\center
\includegraphics[scale=0.6]{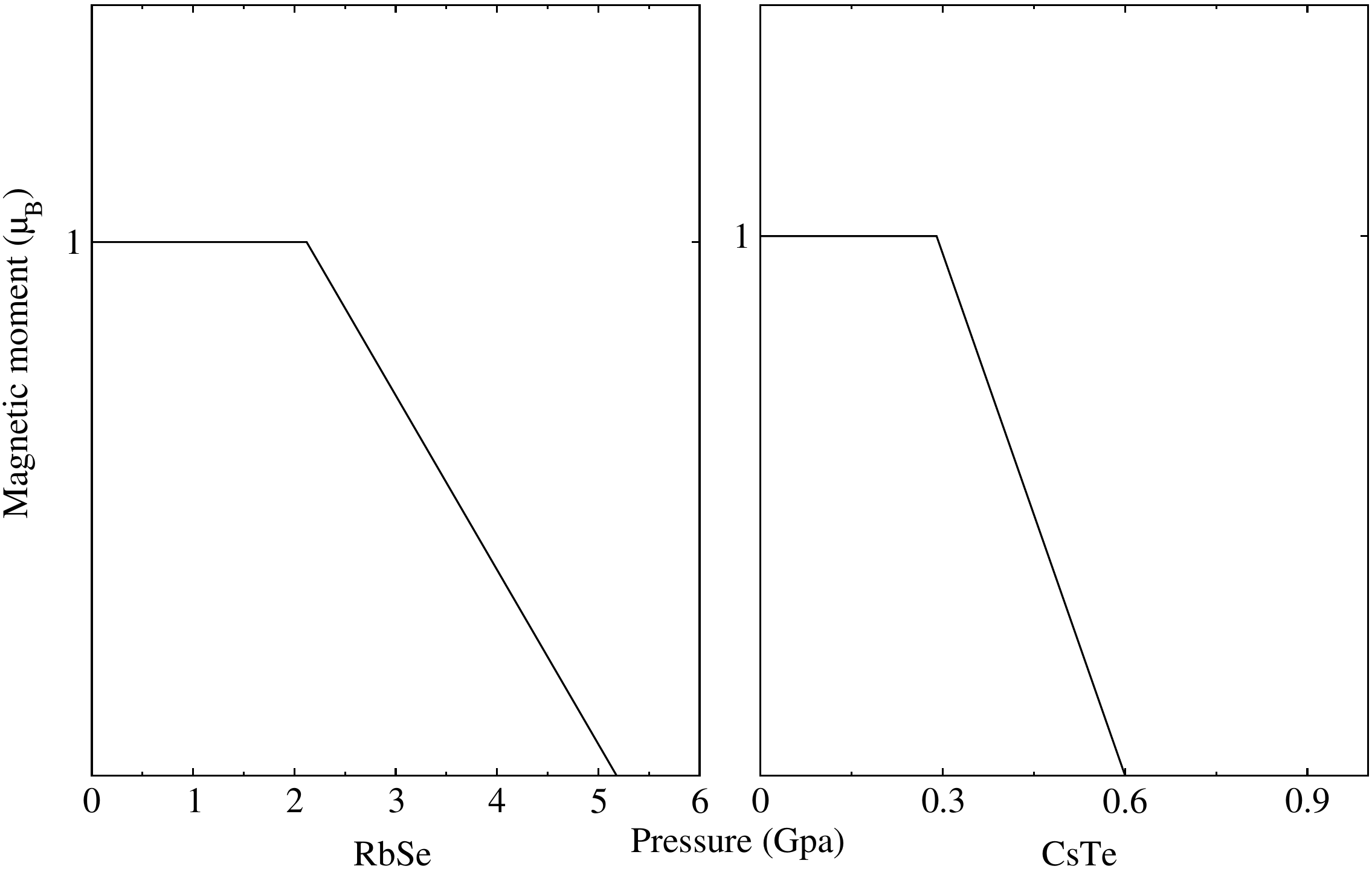}
\caption{The changes of magnetic moments of RbSe and CsTe under pressure, zero pressure is in correspondence to equilibrium lattice constants. }
\label{}
\end{figure}

\end{document}